\documentclass[nohyperref]{article}

\usepackage{microtype}
\usepackage{graphicx}
\usepackage{subfigure}
\usepackage{booktabs} % for professional tables

\usepackage{hyperref}

\usepackage{multicol}

\usepackage[accepted]{icml2022}

\usepackage{amsmath}
\usepackage{amssymb}
\usepackage{mathtools}
\usepackage{amsthm}

\usepackage[capitalize,noabbrev]{cleveref}

\usepackage{soul, float}  %romit

\theoremstyle{plain}

\theoremstyle{definition}

\theoremstyle{remark}

\usepackage[textsize=tiny]{todonotes}

\icmltitlerunning{Learning to Separate Voices by Spatial Regions}

\usepackage{multicol}



\begin{document}

\twocolumn[
% \icmltitle{Self-supervised Region-based Binaural Speech
% Separation}
% \icmltitle{From Source Separation to Self-Supervised Region-based Source Separation}
% \icmltitle{From Generic to Self-Supervised Personalized Source Separation}
% \icmltitle{Trading Audio Source Separation for Audio Region Separation}
\icmltitle{Learning to Separate Voices by Spatial Regions}



\icmlsetsymbol{equal}{*}

\begin{icmlauthorlist}
\icmlauthor{Zhongweiyang Xu}{yyy}
\icmlauthor{Romit Roy Choudhury}{yyy}
% \icmlauthor{Firstname3 Lastname3}{comp}
\end{icmlauthorlist}

\icmlaffiliation{yyy}{Department of Electrical and Computer Engineering, University of Illinois Urbana-Champaign, Illinois, US}

\icmlcorrespondingauthor{Zhongweiyang Xu}{zx21@illinois.edu}
\icmlcorrespondingauthor{Romit Roy Choudhury}{croy@illinois.edu}

\icmlkeywords{Machine Learning, ICML}

\vskip 0.3in
]



\printAffiliationsAndNotice{}  % leave blank if no need to mention equal contribution

\begin{abstract}
We consider the problem of audio voice separation for binaural applications, such as earphones and hearing aids.
While today's neural networks perform remarkably well (separating $4+$ sources with $2$ microphones) they assume a known or fixed maximum number of sources, $K$.
Moreover, today's models are trained in a supervised manner, using training data synthesized from generic sources, environments, and human head shapes. 

% generic audio sources, generic room impulse responses (RIR), and a generic human head-shape (HRTF).
% These limit generalizability to specific users and environments. 

This paper intends to relax both these constraints at the expense of a slight alteration in the problem definition.
We observe that, when a received mixture contains too many sources, it is still helpful to separate them by region, i.e., isolating signal mixtures from each conical sector around the user's head.
% it is still helpful to separate signal mixtures, where each mixture contains signals from a specific region (say the cone in front of the user's face).
This requires learning the fine-grained spatial properties of each region, including the signal distortions imposed by a person's head. 
We propose a two-stage self-supervised framework in which overheard voices from earphones are pre-processed to extract relatively clean {\em personalized} signals, which are then used to train a region-wise separation model.
% into {\em personalized} region-based mixtures, which are then used to train a separation model.
% which in turn help to learn the personalized, fine-grained spatial properties of each region.
Results show promising performance, underscoring the importance of personalization over a generic supervised approach. 
(audio samples available at our project website\footnote{\url{https://uiuc-earable-computing.github.io/binaural/}}).
We believe this result could help real-world applications in selective hearing, noise cancellation, and audio augmented reality.

\vspace{-0.2in}
\end{abstract}

\section{Introduction}
Audio source separation research \cite{fasnet, gu2019endtoend, gu2020enhancing,jenrungrot2020cone} has focused extensively on separating sources from a generic microphone array (e.g., a table-top teleconference system, robots, cars, etc.).
When these microphones are on ``earable’' devices, such as hearing aids and earphones, new opportunities emerge.
% These opportunities are being explored in recent years.
In particular, the human face/ears/head alter the arriving audio signals in sophisticated ways, ultimately helping the brain infer important spatial attributes of the signal \cite{spatial_hearing}.
% which alters the signal that enters the ear, and ultimately helps the brain infer important attributes of the signal.
Importantly, this {\em head-related transfer function} (HRTF) is different across users, and harnessing this personalized filter remains a rich area of exploration in various fields of science \cite{person_hrtf, zhijian}.

\begin{figure}[!h]
\vspace{-0.1in}
  \centering
\includegraphics[width=3in]{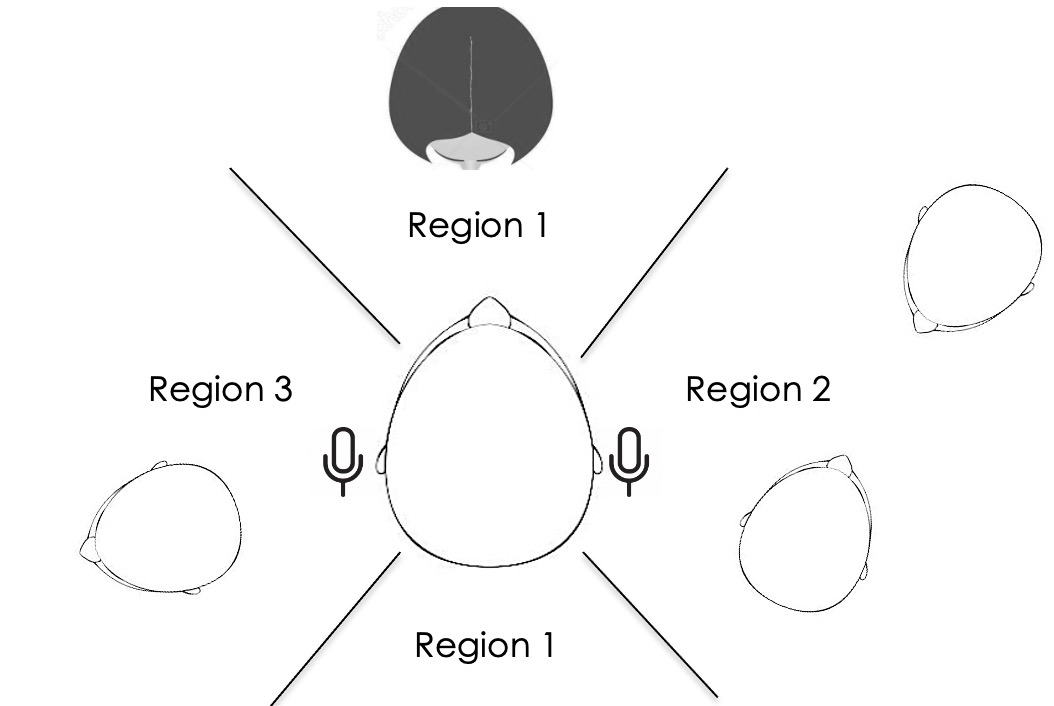}
  \vspace{-0.1in}
  \caption{K=4 sources in R=3 spatial regions (or cones). The front and back cones identified as the same region due to spatial aliasing (i.e., signals arriving from the front and back cones can produce the same time difference at the ears).}
\vspace{-0.05in}
  \label{fig:region}
\end{figure}

This paper aims to explore the potential benefits of personal HRTFs in binaural source separation, such as for hearing aids, earphones, or glasses.
Our hypothesis is that the personal HRTF encodes considerable spatial diversity and this diversity can aid source separation compared to a baseline with generic HRTFs. 
Of course, the spatial diversity may not be enough to separate two sources arriving from nearby angles; in fact even typical humans cannot achieve more than $20^\circ$ angular resolution \cite{zhijian}.
However, if the diversity can separate sources by broad angular regions (e.g., isolate one mixture per region shown in Figure \ref{fig:region}), various applications may benefit from it.
A user’s hearing aid, for example, could help isolate a target voice in front of him from all other voices in other regions.
Even if two voices arrive from the front, separating that two-voice mixture from all other mixtures could still be beneficial to noise cancellation, augmented reality, and other applications.

Separating voices by region has a second benefit.
If the voices can be separated into $R$ mixtures (with one mixture per region), the separation algorithm does not need to know the number of sources, $K$, in the recorded signal.
This offers an important relaxation from past work \cite{2019, luo2020dualpath, sagrnn, subakan2021attention}. 
Specifically, although recent deep learning models have performed remarkably well in source separation (separating $K$=$4+$ sources from $2$ microphone recordings), they have assumed a known $K$ or at least a fixed maximum $K$ \cite{nachmani2020voice}. 
Region-based separation obviates the need to pre-specify $K$ since it would always output at most $R$ mixtures.
Of course, the tradeoff is that the output signals may not be isolated voices, rather they would be mixtures of voices if more than one are located in the same region. 
This paper makes this compromise to gain from HRTF-based personalization and immunity from $K$.

The difficulty in harnessing the personalized HRTF lies in extracting clean binaural signals for any given individual. 
Clearly, a user Alice has abundant opportunities to record ambient signals on her own earphones, but any such binaural recording would likely be polluted and this polluted data would need separation. 
% Thus, recent work \cite{binaural, sagrnn} have used generic data (i.e., anechoic source signals filtered with generic HRTF databases) and used this dataset to train a binaural separation model (a supervised approach).
Thus, recent work \cite{binaural, sagrnn} have used generic data (i.e., anechoic source signals filtered with HRTF databases) and used this clean dataset to train a general binaural separation model (a supervised approach).

To gain from HRTF personalization, one potential approach would be to require the user to record sounds in a quiet room, from many different directions.
This would help train a personalized supervised model, although at the expense of significant user effort.

Our idea is to utilize Alice’s own recordings (from everyday scenarios) and {\em opportunistically} extract out relatively clean signals (whenever possible).
We propose a pre-processing module that uses spectral and spatial techniques to identify {\em when reliable separation is viable}; otherwise, the signal segment is discarded.
Our algorithm relies on a Gaussian mixture model (GMM) to identify when a signal is relatively clean (i.e., high SINR), versus situations where signal-clusters have merged deceptively to appear as one signal.
Moreover, given earphone microphones are separated by a relatively large distance (diameter of human head), the algorithm must also cope with heavy spatial aliasing at higher frequencies.

The output of our pre-processing module is expected to yield relatively clean sources that embed the user’s personalized HRTF.
Using these sources, we synthesize region-wise voice mixtures, and then train a neural network-based separation model.
Since the reference signals are synthesized from Alice’s own recorded signals, {\em the training is self-supervised}.
% We show that even though the reference signals are not accurate, the region-wise separation model can still learn to separate source mixtures effectively.
% Results substantively outperform state of the art methods in which generic HRTFs are used for separation. \hl{[add details.]}
We show that even though the reference signals are not perfectly clean, the region-wise separation model can still learn to separate source mixtures effectively.
Results from our self-trained model outperforms supervised models trained with generic HRTFs by $2$+$dB$.
Our model is not data-hungry and can achieve voice separation without requiring any knowledge of $K$.

% Results from our self-trained model outperforms supervised models trained with \hl{generic HRTFs by $2$+$dB$, while falling $\approx4dB$ short of the best possible model (where both the reference signals and the personal HRTF are accurately known).}
% Results substantively outperform the same model trained with an HRTF database \hl{[add details.]}
% The separation accuracy is even close to the supervised method — the upper bound — in which the reference signals and the personal HRTF are accurately known.

Our contributions are: 
(1) recognizing the combined gain from personalization, self-supervision, and relaxed assumptions on $K$ in exchange for region-wise source separation.
(2) a proposed pre-processing module and a network architecture that realize this gain, and
% (2) an opportunistic voice separation method that extracts \hl{adequately clean} voice sources, and trains a self-supervised model using the extracted dataset, and
(3) extensive comparisons that show how spatial cues (embedded in personalized HRTFs) can play a crucial role in source separation.
% offer consistent gains, without burdening the user with data collection or requiring any knowledge of the number of sources, $K$.

% Our contributions are: (1) an opportunistic voice separation method that extracts \hl{adequately clean} voice sources, (2) a self-supervised model that trains on the extracted dataset, and (3) extensive comparisons that show spatial cues from personalized HRTFs offer consistent gains, without requiring any knowledge of the number of sources, $K$.

% demonstrating that spatial cues from personalized HRTFs can offer sizeable gains, without requiring any knowledge of the number of sources, $K$.

% demonstrating that personalized HRTF in binaural signals embed valuable spatial cues to enable region-wise source separation, without requiring any knowledge of the number of sources, $K$.
% Our model appreciably outperforms competitive schemes that use general HRTFs or attempt classical source separation.

\section{Formulation and Baseline}
\subsection{Problem Formulation} % not sure if I should have this section or not
% We formulate the voice separation problem as follows. 
Consider two microphones at the human ears that hear multiple ambient voices from all around the head.
Assume space is partitioned into $R$ regions, and the signals from each region $i$ form a mixture $y_i$.
These per-region mixtures can be modeled at the left and right microphones as:
\begin{equation}\label{eq1}
    y_i^l = \sum_{j=1}^{N_i} h_{ij}^l \ast s_{ij} \;\;\;\;\;\;\;\;\;\;
    y_i^r = \sum_{j=1}^{N_i} h_{ij}^r \ast s_{ij}
\end{equation}
Here $s_{ij} \in \mathbb{R}^{1 \times T}$ is the $j^{th}$ signal in the $i^{th}$ region; $h_{ij}^l$ and $h_{ij}^r$ are the corresponding head-related impulse response (HRIR) for the direction from which source $s_{ij}$ arrives.
The $\ast$ denotes the convolution operation and 
$N_i$ denotes the number of sources in the $i^{th}$ region. 
Then, the recorded mixture at at each microphone would be a summation over all region-based mixtures as follows:
\begin{equation}\label{eq2}
    m^l = \sum_{i=1}^{R} y_i^l \;\;\;\;\;\;\;\;\;\;
    m^r = \sum_{i=1}^{R} y_i^r
\end{equation}
Here $m^l$ and $m^r$ are the left and right microphone recordings.
The goal of region-based separation is to estimate $y_i^l$ and $y_i^r$ from $m^l$ and $m^r$, for all $i \in [1,R]$.

% Suppose there are $R$ regions around the human head. 
% One possibility of the region setup is shown on figure 1. For each region i, there exists $y_i$ consisting of all sounds inside that region. For i from 1 to $R$, we have:
% % \begin{flalign*}
% %     \begin{aligned}
% %     y_i^l = \sum_{j=1}^{N_i} h_{ij}^l \ast s_{ij} \;\;\;\;\;\;\;\;\;\;
% %     y_i^r = \sum_{j=1}^{N_i} h_{ij}^r \ast s_{ij}\\
% %     \end{aligned}
% % \end{flalign*}
% \begin{equation}\label{eq1}
%     y_i^l = \sum_{j=1}^{N_i} h_{ij}^l \ast s_{ij} \;\;\;\;\;\;\;\;\;\;
%     y_i^r = \sum_{j=1}^{N_i} h_{ij}^r \ast s_{ij}\\
% \end{equation}
% Here $s_{ij} \in R^{1xT}$ is the $j_{th}$ signal in the $i_{th}$ region. $h_{ij}^l$ and $h_{ij}^r$ is the corresponding head-related impulse response regarding the direction of the source $s_{ij}$. $N_i$ represents the number of sources in the $i_th$ region The $\ast$ represents convolution here. Then the mixture recording would a summation of all region mixtures:

\subsection{Supervised Separation as Baseline}
% For the region-based voice separation task, we first propose a supervised training approach. 
% This supervised model will be trained with clean but generic data and serve as a baseline for comparison.
% In contrast, our self-supervised architecture will lack clean training data but will benefit by learning personalized room and head-related filters.
% Of course, the upper bound will be achieved by training a supervised model with clean personalized data.
% Our results will report on all these architectures.

For a supervised approach, the binaural mixtures $m_l$ and $m_r$ are fed into a separation model $f_{\theta}$ with parameters $\theta$. 
The model predicts $R$ binaural sounds which corresponds to $R$ regions: 
$$\hat{y}_1^l,\;\hat{y}_1^r,\; ~~ \hat{y}_2^l,\;\hat{y}_2^r,\; ~~ ...\;, \hat{y}_R^l,\;\hat{y}_R^r = f_{\theta}(m_l, m_r)$$
To optimize the model parameters $\theta$, the loss function of our system contains two parts: {\em active loss} and {\em inactive loss}. 
The active loss is for the regions that have voices --- we want each region's output to contain all the sources inside that region.
The inactive loss is for the regions those contain no active voices --- we want these regions to output an empty source.
Thus we adopt the loss function in \cite{fuss} but without permutation invariant training (PIT). 
Specifically, assume reference region mixtures we want to learn are $y_i^l,y_i^r,i\in[1, R]$, ordered so that first $M$ reference region mixtures are active. Then, the loss function is:
\begin{flalign*}
    \begin{aligned}
    Loss &= \sum_{i=1}^{M} [L_{\text{SNR}}(y_i^l, \hat{y_i}^l)+L_{\text{SNR}}(y_i^r, \hat{y_i}^r)]\\
    &+ \sum_{i=M+1}^{R} [L_{\text{inactive}}(m^l, \hat{y_i}^l)+L_{\text{inactive}}(m^r, \hat{y_i}^r)]\\
    % y_i^l = \sum_{j=1}^{N_i} h_{ij}^l \ast s_{ij} \;\;\;\;\;
    % y_i^r = \sum_{j=1}^{N_i} h_{ij}^r \ast s_{ij}\\
    \end{aligned}
\end{flalign*}
The $L_{\text{SNR}}$ loss in the equation below is the negative SNR loss with a constant $\tau$=$10^{-\text{SNR}_{\text{max}}/10},\;\text{SNR}_{\text{max}}$=$30dB$. 
\begin{flalign*}
    \begin{aligned}
    L_{\text{SNR}}(y,\hat{y}) = 10log_{10}(||y-\hat{y}||^2+\tau||y||^2)
    \end{aligned}
\end{flalign*}
The constant $\tau$ is to assign a $30dB$ maximum SNR to prevent the network from optimizing for one single source. 
The detailed reasoning is explained clearly in \cite{mixit,fuss}.

The $L_{inactive}$ is to enforce the network to output empty source for inactive regions. Thus the inactive loss is: 
\begin{flalign*}
    \begin{aligned}
    L_{\text{inactive}}(x,\hat{y}) = 10log_{10}(||\hat{y}||^2+\tau||x||^2)
    \end{aligned}
\end{flalign*}
For supervised training, we generate training data by convolving voice sources with human head-related impulse responses (HRIR)\footnote{HRIR is the time domain representation of an HRTF. The HRIR varies as a function of the signal's direction of arrival.}.
% For supervised training, use voice sources, known HRTFs are used to synthesize the mixtures and clean reference signals. 
% Assume we want the model to provide separation for a specific person.
We consider two cases: 
First, we assume we know the person's HRTF; we use this filter to create the dataset and train the separation model.
Assume the separation performance is $P_{\text{personal}}$ for this case. 
% First, say we know the person's HRTF and we train the separation model using this specific HRTF only. 
% Assume the separation performance is $p1$ for this case. 
Second, assume we don't know the person's HRTF and we train the separation model using a generic HRTF database. 
Say the separation performance for this person is $P_{\text{general}}$. 
Obviously, $P_{\text{personal}} > P_{\text{general}}$ because the first case is learning the person's personalized HRTF.
However, $P_{\text{personal}}$ is not achievable since the personal HRTF of a given person is not known in practice.
Our goal is to opportunistically learn the personalized HRTF at a region-wise granularity --- we expect that these personal spatial cues, even though self-supervised (hence imperfect), will help outperform $P_{\text{general}}$ and take us close to $P_{\text{personal}}$.

% Our goal is to outperform $P_{\text{general}}$ by leveraging the spatial cues from personalized HRTFs, and strive for $P_{\text{personal}}$ by learning the personal HRTF (at a region-wise granularity) from the self-recorded binaural signal. 
% In other words, we ideally want: $P_{\text{personal}} \approx P_{\text{personal}} > P_{\text{general}}$.
% This enables our self-supervised separation model.

% Therefore, we ask, is it possible to outperform $p2$ and get close to $p1$ without prior knowledge of the specific person's HRTF.
% This motivates our self-supervised training pipeline, proposed next.

% knowledge of personal HRTFs, or clean training data.
% by learning the personalized HRTF on the fly, in a self-supervised manner.
% Given this gap $p1 - p2$, we ask is if it is possible to achieve a separation performance $p3$ that is closer to $p1$ than $p2$, but without 
% closer to $p1$ than $p2$, but still without using the specific person's HRTF. Thus, we propose a self supervised training pipeline.

\section{Two Stage Model}
The first stage aims to accept binaural recordings from a user's ear-device and output relatively clean voice sources along with their directions of arrival (DoA).
The output sources should not be contaminated too much so that they preserve the personal HRTF (naturally embedded in the signals).
The challenge lies in identifying and eliminating deceptive mixtures that appear as single or two sources, or when two sources appear separable but have corrupt spatial cues.

In stage 2, we use these sources and their DoAs to synthesize larger mixtures of many sources {\em per region}.
This region-wise mixture-dataset is then used to train our separation model.
The output sources from stage 1 serve as reference signals for our loss function, thereby self-training the model.
We elaborate on the two stages next.

\subsection{Stage 1: Reliable Source Extraction}
\begin{figure}[!h]
  \centering
  \includegraphics[width=3.4in]{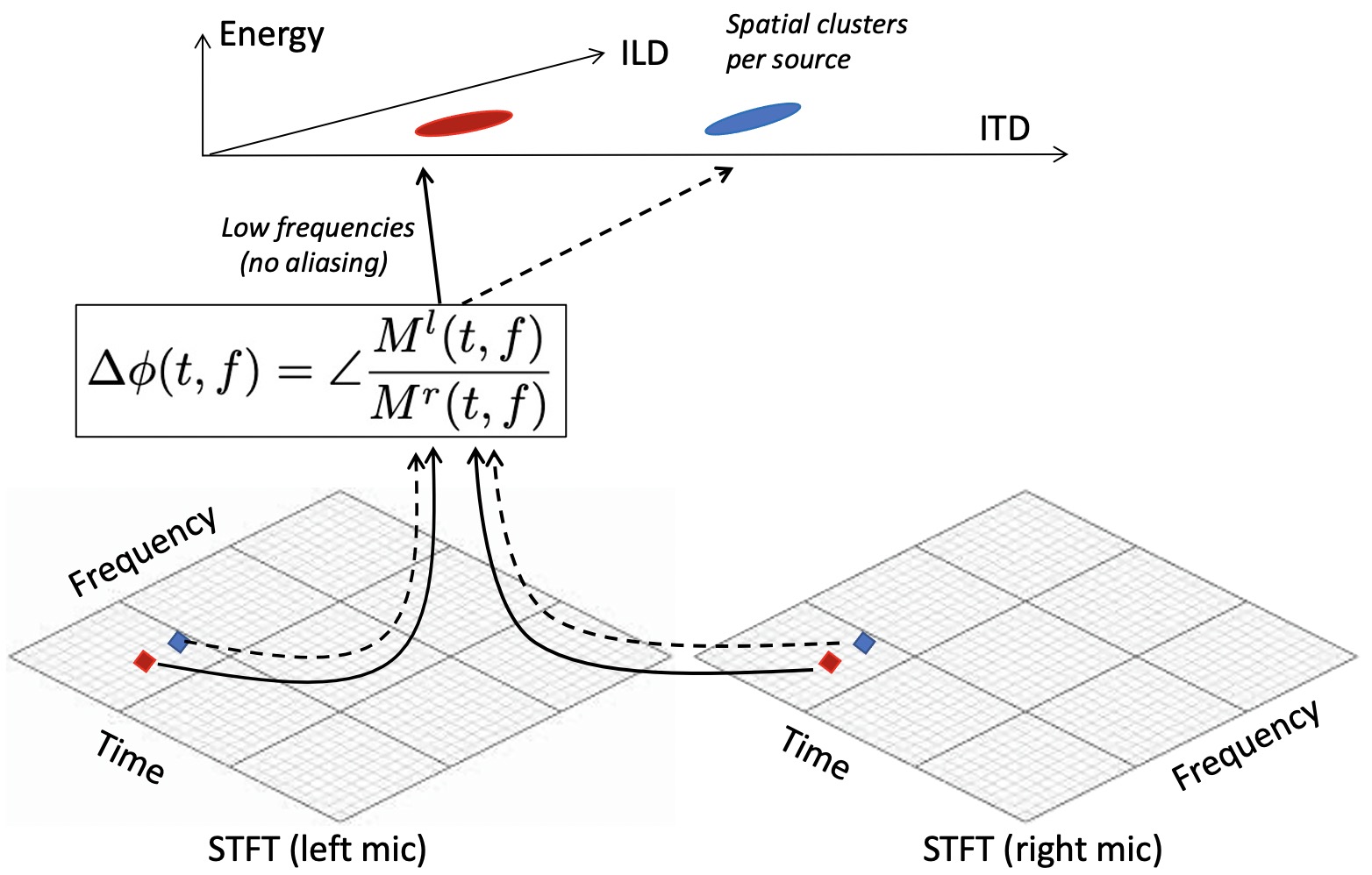}
  \vspace{-0.2in}
  \caption{From STFT to clustering on the ITD--ILD space.}
  \vspace{-0.1in}
  \label{fig:cluster}
\end{figure}

We intend to spatially cluster the two-microphone recordings, but such techniques are not without limitations.
Today, state-of-the-art spatial clustering exploits the inter-microphone time difference (ITD) and inter-microphone level difference (ILD) as the key spatial features for clustering \cite{duet, mandel2007algorithm, mandel2007localization, mandel2008, mandel2009model}.
Briefly, if $M^l(t, f)$ and $M^r(t, f)$ denote the STFT of $m^l$ and $m^r$, we can calculate the inter-microphone phase difference $\Delta\phi(t,f)$ for each time--frequency ($t$--$f$) bin:
\begin{equation}\label{eq1}
    \Delta\phi(t, f) = \angle{\frac{M^l(t, f)}{M^r(t, f)}} \;\;\;\;\;\;\
\end{equation}
Assuming the two microphones are sufficiently close to avoid spatial aliasing, the ITD for each $t$--$f$ bin can be directly estimated from $\Delta\phi(t, f)$ as $\text{ITD}(t, f) = \frac{\Delta\phi(t, f)}{2\pi{f}}$.
Figure \ref{fig:cluster} visualizes this where $\Delta\phi(t, f)$ is computed from two low frequency $t$-$f$ bins, and mapped to the ITD axis.

Similarly, the amplitude-level difference ILD can be computed as:
\vspace{-0.15in}
\begin{equation}\label{eq1}
    \text{ILD}(t, f) = 20log_{10}\frac{|M^l(t, f)|}{|M^r(t, f)|}
\end{equation}

Assuming the ILD only varies with direction (and not across frequency), it is possible to spatially cluster on the ITD+ILD dimensions.
Each cluster maps back to the $t$--$f$ masks on the STFT, ultimately achieving decent spatial separation of the two (red and blue) sources.

% This leads to $t$--$f$ masks (on the STFT) that ultimately yields decent spatial separation.

% and assuming the signals are in low frequencies, the ILD only varies with direction and not frequency.

% % and assuming the signals source are in the near field, \hl{the ILD exhibits spatial variation.} 
% Thus, clustering on the ITD+ILD dimensions lead to $t$--$f$ masks (on the STFT), that ultimately yield descent spatial separation.

Unfortunately, limitations emerge even if all the above assumptions hold: 
(1) Clustering assumes each $t$--$f$ bin could only contain one single source -- \cite{duet} calls this assumption W-Disjoint Orthogonality (W-DO). 
This approximately holds with $K=2$ or $3$ voice sources; with more sources, signals ``collide'' in $t$--$f$ bins.
(2) The second problem is that if two sources arrive from nearby angles, their spatial features blend into a single cluster, making separation impossible. 
% For region-based separation, we will be able to utilize such blending of clusters since it gives us a mixture from a specific region.

\begin{figure}[!h]
  \centering
  \includegraphics[width=3.4in]{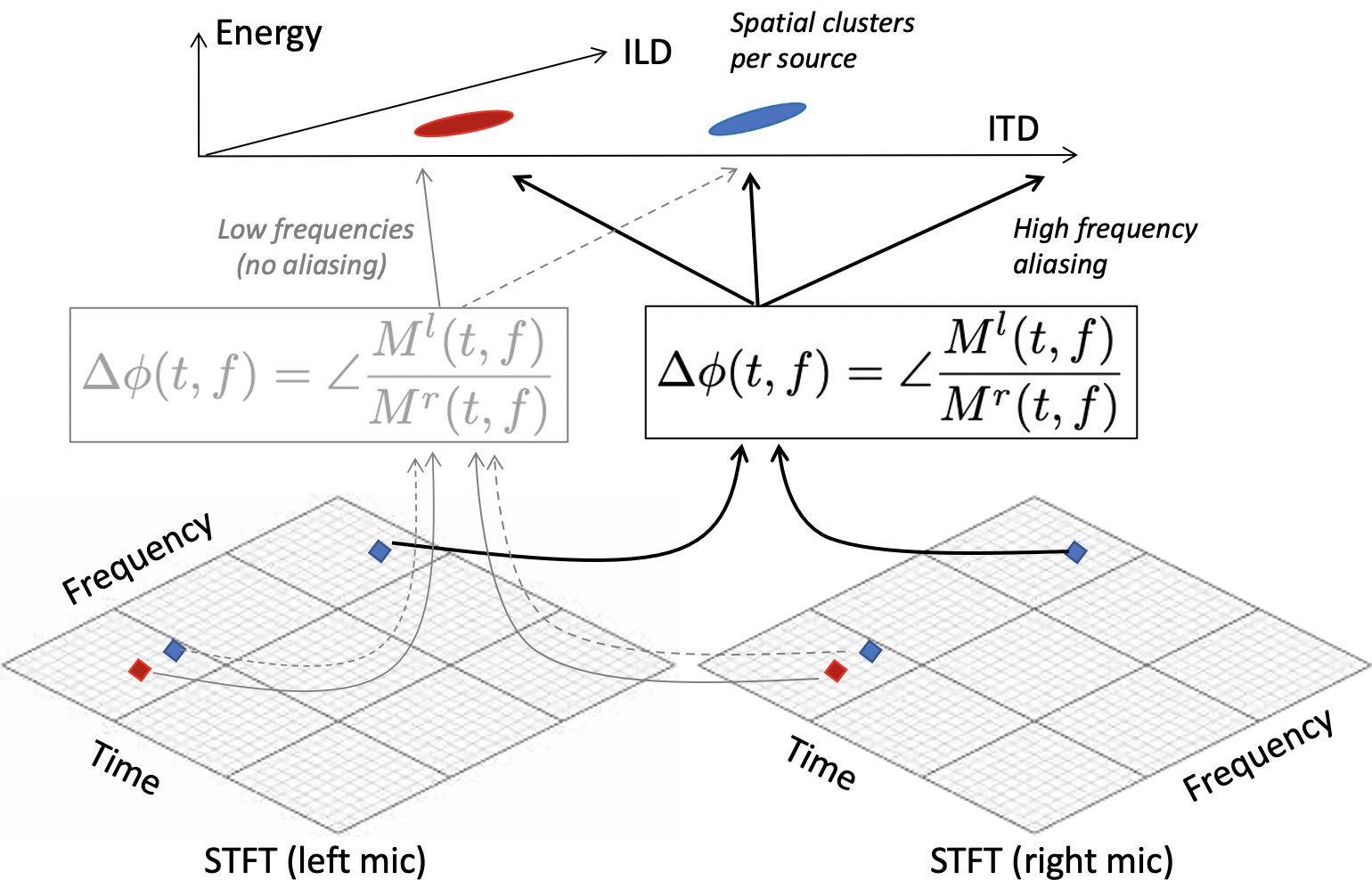}
  \vspace{-0.2in}
  \caption{Higher frequencies produce spatial aliasing (i.e., due to smaller wavelengths, the measured phase difference can translate to many possible time differences or ITDs). The example blue signal in the high-frequency $t$--$f$ bin shows $3$ possible ITDs , which affects the separation of red and blue source signals.}
  \vspace{-0.1in}
  \label{fig:alias_cluster}
\end{figure}

In non-ideal cases, i.e., when the assumptions do not hold, additional issues emerge.
(3) Since human heads are relatively large in comparison to the wavelength at higher voice frequencies, spatial aliasing becomes a problem. 
If $\Delta\tau_{\text{max}}$ is the maximum inter-microphone time difference (between the two ears), the minimum frequency for possible aliasing is: $f_{\text{aliasing}} = \frac{1}{2\times\Delta\tau_{\text{max}}}$.
For average human head shapes, $f_{\text{aliasing}} \approx 1000 \textit{Hz}$, implying that more than $87\%$ of the frequencies get aliased (Figure \ref{fig:alias_cluster} shows the multiple ambiguous ITDs due to spatial aliasing from the high frequency $t$--$f$ bins).
(4) Finally, the human HRTF is frequency selective, hence the ILD also varies per-frequency; modeling this variation is hard since it is unique to each individual.
To mitigate these $4$ problems, we design a {\em selective spatial clustering} algorithm, discussed next.

\subsection*{Selective Spatial Clustering}
Algorithm \ref{alg:selective} presents the pseudo code; 
we explain the key steps below.

\textbf{Step 1:} 
We conservatively estimate $f_{aliasing}$ (based on maximum possible human head size) and use the unaliased frequency bins to estimate ITD.
We cluster on ITD and look for $1$ peak or $2$ adequately separated peaks. 
% clean peaks.

\textbf{Step 2:} 
A single peak indicates either a single source, or multiple (angularly) nearby sources that have merged (in ITD) to become a single peak.
We fit a Gaussian on this peak and accept the peak if the estimated variance is less than a threshold $\sigma_{th}^2$.

\begin{figure}[!h]
\vspace{-0.1in}
  \centering
\includegraphics[width=3.2in, keepaspectratio]{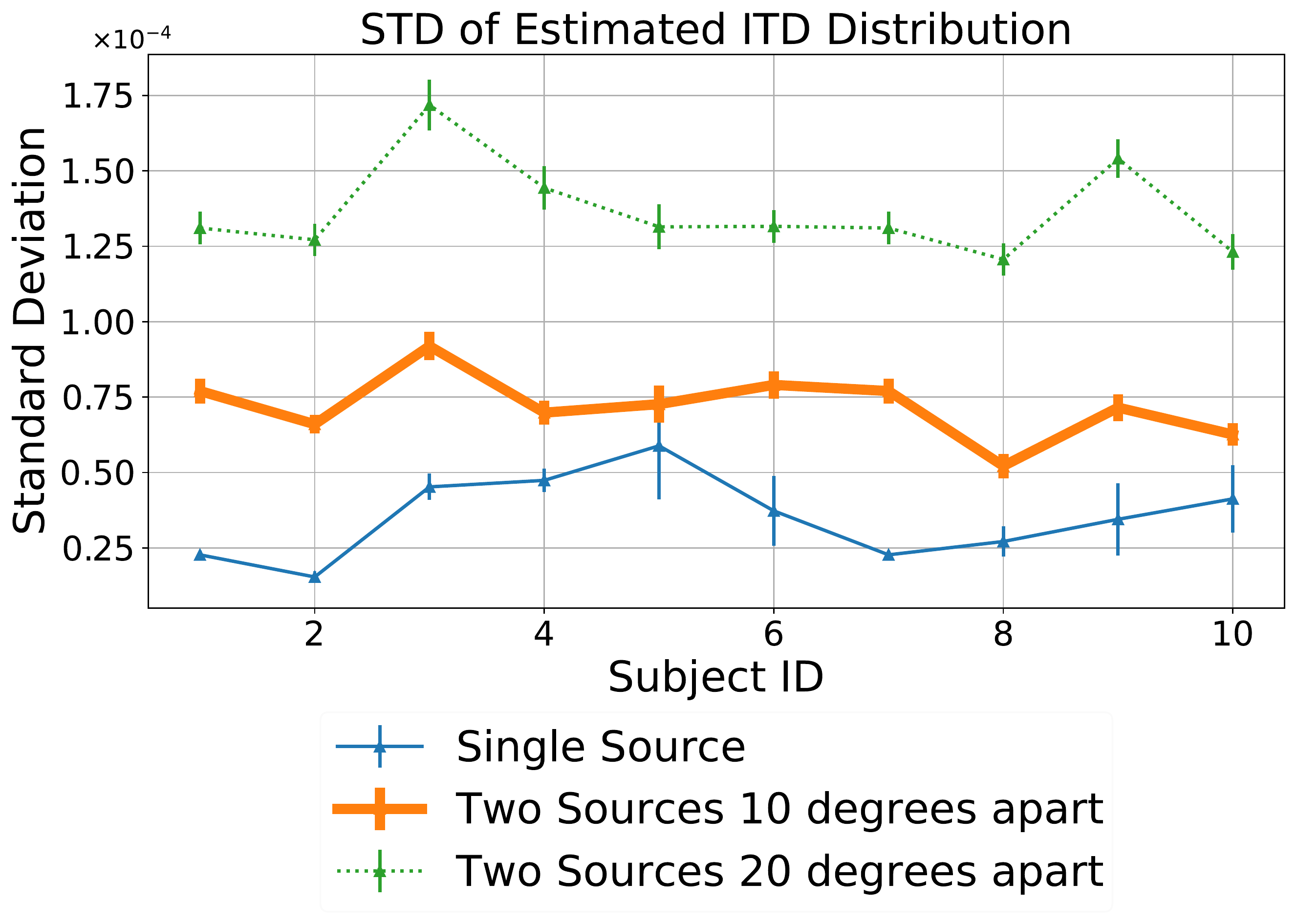}
  \vspace{-0.1in}
  \caption{Standard deviation of estimated ITDs for single clean source and 2-source mixtures with $10$ and $20$ degree apart. The wide gap between $10$ and $20$ degrees allows for reliable separation.}
\vspace{-0.15in}
  \label{fig:sigma}
\end{figure}

We show that the standard deviation (STD) of the ITD distribution is a robust indicator of whether voice directions are angularly nearby or separated.
% are arriving from one single direction or if they are separated by certain angles. 
Figure \ref{fig:sigma} shows a clear STD gap between a single source and 2-source mixtures when the $2$ sources are $20^\circ$ apart.
This guides our choice of $\sigma_{th}^2$.
% We show that the standard deviation of the ITD distribution is a robust indicator of whether voices are arriving from one single direction or if they are separated by some certain angles. Figure \ref{fig:sigma} shows that there exists an obvious bar between single source's STD and 2-source mixture's STD when the two sources are 20 degrees apart.

% \begin{figure}[!h]
% \vspace{-0.1in}
%   \centering
% %   \includegraphics[width=2in]{images/region_setup.pdf}
% \includegraphics[width=3.2in, keepaspectratio]{images/sigma.pdf}
%   \vspace{-0.2in}
%   \caption{Standard deviation of estimated ITDs for single clean source and 2-source mixtures with 10 and 20 degree apart.}
% \vspace{-0.15in}
%   \label{fig:sigma}
% \end{figure}
The time-frequency mask corresponding to this ITD peak gives us one source or one mixture, from a specific region.
We add this source to our database.

\textbf{Step 3:} 
Similarly, $2$ peaks indicate $2$ sources or $2$ mixtures, but we need them to be sufficiently separated in ITD to gain confidence that they have not mutually contaminated each other.
For this, we fit the ITDs to a 2-component {\em gaussian mixture model} (GMM) and check if the variances are less than $\sigma_{th}^2$, 
and their means differ more than $\Delta\tau_{min}$.
If the fitted Gaussians satisfy none of these (conservative) conditions, we deem the sound segment unsuitable for separation and discard it. 
Otherwise, we proceed to further separation.

\textbf{Step 4:} 
Given audio frequencies far exceed $f_{aliasing}$, we need to cluster on high frequency bins (and cope with ITD aliasing).
Past work \cite{mandel2007algorithm, mandel2007localization, mandel2008, mandel2009model} shows that inter-microphone phase differences (IPD) are highly noisy at these frequencies, but ILDs are helpful due to greater frequency-sensitivity. 
Motivated by this, we aim to estimate ILDs for the two spatial sources.

\begin{algorithm}[H]
% \begin{breakablealgorithm}
\caption{Selective Spatial Separation}
\label{alg:selective}
% \onecolumn 
\begin{algorithmic}
   \STATE {\bfseries Input:} 
        \\\;\;\;\;Real world binaural recording $m^l, m^r$\\
        \;\;\;\;Aliasing frequency threshold $f_{aliasing}$\\
        \;\;\;\;Variance threshold for peak detection $\sigma_{th}^{2}$\\
        \;\;\;\;Dual source ITD difference threshold $\Delta\tau_{min}$\\
        \;\;\;\;Time Domain Dominating Factor $\alpha$
    \STATE {\bfseries \textit{Step 1: Use unaliasing t-f bins to find ITD distribution}} \\
    \STATE $M^l(t, f) \gets STFT(m^l)\;\;\;\;M^r(t, f) \gets STFT(m^r)$
    % \STATE $M^r(t, f) \gets STFT(m^r)$
    \STATE $\Delta\phi(t, f\in(0:f_{aliasing})) \gets \angle{\frac{M^l(t, f\in(0:f_{aliasing})}{M^r(t, f\in(0:f_{aliasing}))}}$
    \STATE $ITDs(t, f\in(0:f_{aliasing})) \gets \frac{\Delta\phi(t, f\in(0:f_{aliasing}))}{2\pi{f}}$
    \STATE {\bfseries \textit{Step 2: Output the binaural recording and the esimated ITD when the ITDs show one single obvious peak}} \\
    % \STATE $\mu^{\ast}, \sigma^{\ast} \gets \underset{\mu, \sigma}{\arg\max}\; \mathcal{N}(ITDs; \mu,\,\sigma^{2})$
    \STATE $\mu^{\ast}, \sigma^{\ast} \gets {\arg\max}\; \mathcal{N}(ITDs; \mu,\,\sigma^{2})$
    \IF{$\sigma^{\ast} < \sigma_{th}$}
    \STATE Return $m^l, m^r, ITD=\mu^\ast$
    \ENDIF
    % \STATE ${\mu_1}^{\ast}, {\sigma_1}^{\ast}, {\mu_2}^{\ast}, {\sigma_2}^{\ast} \gets \underset{\mu_1, \sigma_1, \mu_2, \sigma_2}{\arg\max}\; Gaussian Mixture(ITDs; \mu,\,\sigma^{2})$
    \STATE {\bfseries \textit{Step 3: Discard the recordings when the ITDs does not show two obvious peaks, or the two peaks are too close}} \\
    \STATE Use 2-component Gaussian Mixture Model to fit ITDs
    \STATE Let $\mu_1^\ast, \mu_2^\ast, \sigma_1^\ast, \sigma_2^\ast$ be the optimized means and standard deviations of the two components
    
    \IF{$\sigma_1^\ast > \sigma_{th}$ or $\sigma_2^\ast > \sigma_{th}$ or $|\mu_1^\ast-\mu_2^\ast|<\Delta\tau_{min} $}
    \STATE Discard $m^l, m^r$, Return
    \ENDIF
    \STATE {\bfseries \textit{Step 4: Cluster the non-aliasing frequency bins and generate non-aliasing t-f bin masks}} \\
    \STATE Cluster the t-f bins using the GMM for $f < f_{aliasing}$, get two unaliasing t-f bin masks $mask_1^{u}(t, f), mask_2^{u}(t, f)$ for the two separable spatial sounds\\
    % \STATE {\bfseries \textit{Step 5: Get $E_1(t), E_2(t)$, each source's energy at different time instances at non-aliasing frequencies, and then get $T_1, T_2$, the active time instances for each source. Use $T_1, T_2$ and ILDs to estimate the two sources' ILDs($ILD_1(f), ILD_2(f)$) for aliasing frequency bins. Use the estimated ILDs to cluster alisasing frequency bins}} \\
    \STATE {\bfseries \textit{Step 5: Use each source's dominating time bins to estimate the source's ILD(f) for aliasing frequencies.}} \\
    
    % \COMMENT{Below is estimating which source is active at what time}
    \;\;\;\;$E_1(t)\gets\sum_{f}(mask_1^{u} \times (|M^l(t,f)|^2 + |M^r(t,f)|^2)))$\\
    \;\;\;\;$E_2(t)\gets\sum_{f}(mask_2^{u} \times (|M^l(t,f)|^2 + |M^r(t,f)|^2)))$\\

    \REPEAT
    \STATE$T_1 \gets \{t|\;E_1(t) > \alpha\times{E_2(t)}\}$
    \STATE$T_2 \gets \{t|\;E_2(t) > \alpha\times{E_1(t)}\}$
    \STATE $\alpha \gets 0.9\times\alpha$
    \UNTIL{None of $T_1$ and $T_2$ is empty}
    
    % \STATE {\bfseries \textit{Step 6: Use $T_1, T_2$ and ILDs to estimate the two sources' ILDs($ILD_1(f), ILD_2(f)$) for aliasing frequency bins. Use the estimated ILDs to cluster alisasing frequency bins}} \\
    \STATE $ILDs(t,f)\gets20log_{10}\frac{|M^l(t, f)|}{|M^r(t, f)|};\;\;\;\;f\geq f_{aliasing}$
    \STATE $ILD_1(f)\gets{mean(\{ILDs(t, f)|t\in{T_1},\;f\geq f_{aliasing}\})}$
    \STATE $ILD_2(f)\gets{mean(\{ILDs(t, f)|t\in{T_2},\;f\geq f_{aliasing}\})}$
    \STATE $ILD_{threshold}(f)\gets{\frac{ILD_1(f) + ILD_2(f)}{2}}$
    \STATE Use the $ILD_{threshold}$ to get masks for aliasing frequency t-f masks $mask_1^{a}, mask_1^{a}$ 
    \STATE {\bfseries \textit{Step 7: Apply both masks and return the separated spatial sounds with ITD labels }} \\
    \STATE $mask_1(t, f) \gets concatenate(mask_1^u(t, f), mask_1^a(t, f))$
    \STATE $mask_2(t, f) \gets concatenate(mask_2^u(t, f), mask_2^a(t, f))$
    \STATE $s_1^{l,r} \gets iSTFT(mask_1(t, f)\times{M^{l,r}})$
    \STATE $s_2^{l,r} \gets iSTFT(mask_2(t, f)\times{M^{l,r}})$
    \STATE Return $s_1^{l,r}, s_2^{l,r}, ITD_1=\mu_1^{\ast}, ITD_2=\mu_2^{\ast}$
    
    %(\mu, \sigma) \gets Fit $ITDs(t, f\in(0:f_{aliasing}))$ to Gaussian 

%   \REPEAT
%   \STATE Initialize $noChange = true$.
%   \FOR{$i=1$ {\bfseries to} $m-1$}
%   \IF{$x_i > x_{i+1}$}
%   \STATE Swap $x_i$ and $x_{i+1}$
%   \STATE $noChange = false$
%   \ENDIF
%   \ENDFOR
%   \UNTIL{$noChange$ is $true$}
\end{algorithmic}
% \end{breakablealgorithm}
\end{algorithm}

If only one source was active at time $t$, per-frequency ILD estimation would be easy --- we would record the ILDs for each high frequency bin.
With mixtures of signals, this is problematic. 
However, given source signals are mostly uncorrelated, we expect to find time bins in which only one of the sources dominate.
How can we tell when one source dominates? 

\begin{figure*}[h]
  \includegraphics[width=\textwidth,height=6cm]{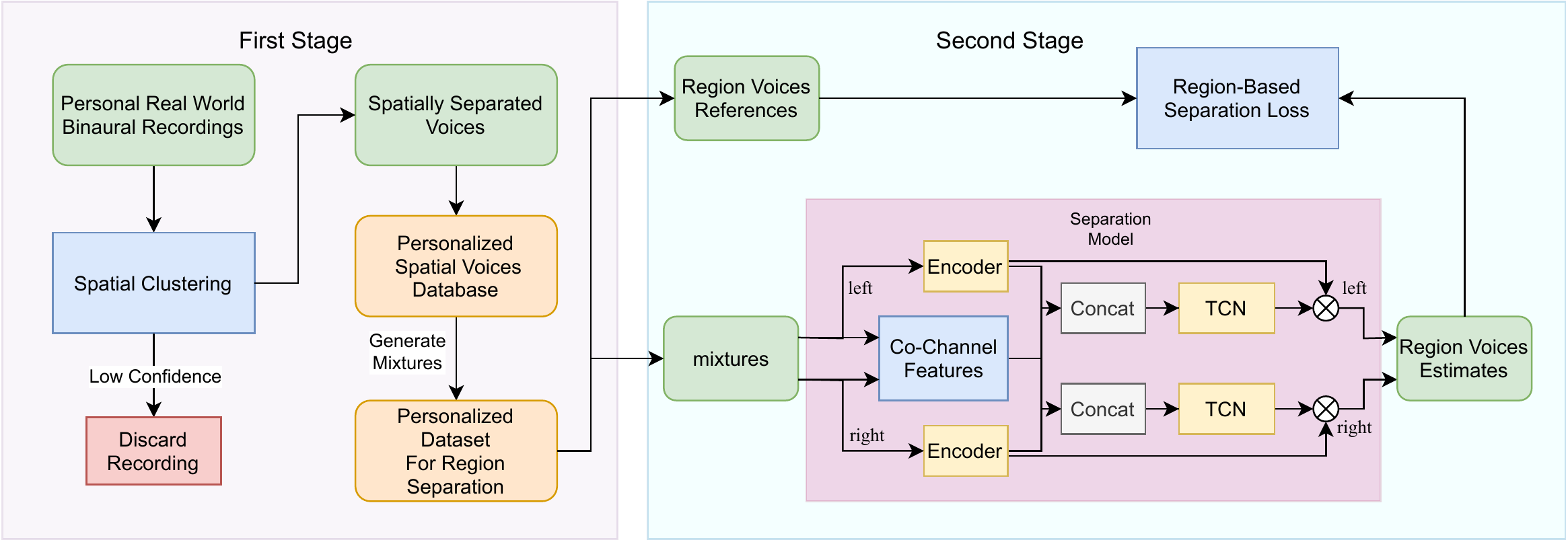}
  \vspace{-0.2in}
  \caption{Two stage pipeline for self-supervised region-based separation using real-world personal recordings. The first stage collects binaural sounds and uses spatial clustering to create a personal database. The second stage uses this database for self-supervised training.}
  \label{fig:pipeline}
\end{figure*}

\textbf{Step 5:} 
We compute the time--frequency masks estimated from the lower (unaliased) frequencies, compute the energy corresponding to each mask, and test if energy $E_1(t)$ exceeds $E_2(t)$ by a factor of $\alpha$.

If source $1$ dominates at certain time instants $T_i$, we compute the mean per-frequency ILD from those time instances.
We perform the same for source $2$.
This yields the per-frequency ILD for each source.

\textbf{Step 6:} 
For all time bins where no source dominates, we compute the ILD for each high frequency bin and compare against the mean ILDs recorded in Step 5.
That frequency bin is assigned to the source whose ILD matches better.
At this point, every $t$-$f$ bin has been assigned a mask.

\textbf{Step 7:} 
The masks are applied and after an inverse STFT, the two signals (or mixtures) are extracted.
The ITDs corresponding to the signals/mixtures are recorded -- this gives us the region from which the signal arrived.
These separated signals/mixtures and their associated regions are entered into a region-wise source database.

% \begin{figure}
%   \centering
%   \includegraphics[width=2in]{images/self_supervised.drawio (2).pdf}
%   \vspace{-0.1in}
%   \caption{The self-supervised training pipeline for personalized region-based voice separation.}
%   \label{fig:speech_production}
% \end{figure}

% \begin{figure}

%   \centering
%   \includegraphics[width=3.2in]{images/nnmodel.jpg}
%   \vspace{-0.1in}
%   \caption{The separation model}
%   \label{fig:nnmodel}
% \end{figure}

\subsection{Stage 2: Self-trained Region-wise Separation Model}
Figure \ref{fig:pipeline} shows our model pipeline -- the output from stage 1 is a database of relatively clean sources (and their DoAs).
Stage 2 uses these sources to synthesize region-wise mixtures and then {\em mixes these mixtures} to create the binaural recordings.
This dataset trains our separation model.

% The database of region-wise source signals/mixtures allows synthesis of any kind of binaural mixtures with any number of sources. 
% Importantly, these sources embed the user's room impulse responses (RIR) and her personal HRTF, obviating any ground truth around the user's HRTF or RIR distribution.
% Hence, we synthesize mixtures of many sources across many regions and create a personalized dataset for self-supervised training.

Our neural network model for region-based voice separation is the feature concatenation TasNet, derived from \cite{binaural}. Single-channel TasNet contains three modules: encoder, Temportal Convolutional Network (TCN), and decoder. The linear encoder is a list of kernels to transform the time domain signal to an STFT-like 2-d representation. This representation is fed into the TCN module to predict a real mask for each source. After the masks are applied to the representation, the linear decoder transforms the masked representations back to time domain.

The feature concatenation TasNet uses $\cos(\text{ITD})$, $\sin(\text{ITD})$, and $\text{ILD}$ of all time-frequency bins as co-channel features. 
The co-channel features are then concatenated with the encoder output of left or right recordings (in the channel dimension) to generate a new representation for both the left and right channel's mixture. Then representations are fed into TCN to yield left/right separation masks for all sources, respectively. The remainder is same as single-channel TasNet, except that our architecture is designed to output sources for both left and right channels.

We expect our model to learn the spatial cues (from the personalized HRTFs) embedded in the binaural recordings.
We expect that advantages from the personalized spatial information will outweigh the disadvantage of {\em partially-clean} reference signals, outperforming supervised training models that use generic HRTFs.
Finally, our model makes no assumptions on the number of sources, $K$.

\section{Experiments and Evaluation}

% \begin{figure}
% [!h]
%   \centering
%   \includegraphics[width=2in]{images/binaural_model.drawio.pdf}
%   \vspace{-0.1in}
%   \caption{Binaural Feature Concatenation Network Achitecture}
%   \label{fig:speech_production}
% \end{figure}

% \subsection{Self-Supervised and Supervised Training}
% \begin{figure*}
%   \includegraphics[width=\textwidth,height=4cm]{images/binaural_model.drawio (2).pdf}
%   \caption{Binaural Separation Network}
%   \label{fig:network}
% \end{figure*}

To configure the feature concatenation TasNet, we set $N=512, L=32, B=128, Sc=128, P=3, X=8, R=3$, following the convention in \cite{2019}. To calculate the co-channel features $\cos(\text{ITD})$, $\sin(\text{ITD})$, and $\text{ILD}$, we use $256$-bin STFT with hop size $16$ to make sure the STFT can be aligned with the encoder output. Hanning window is applied when calculating the STFT. 
The model is trained on $4$ 1080ti GPUs using the ADAM optimizer with batch size $4$. 
The learning rate is set to be $10^{-3}$.

\subsection{Region-Based Supervised Training}

{\bf HRTF dataset:} 
For supervised region-based separation, we use the CIPIC HRTF database \cite{CIPIC}.
The CIPIC HRTF database contains real-world recorded Head Related Impulse Responses (HRIR) for $45$ subjects, with $50$ different azimuths and $25$ different elevations, at roughly $5$ degrees of angular resolution. 
For our experiments, we only use the horizontal plane with $50$ different azimuth angles, divided into three regions, as shown in Figure \ref{fig:region}. 
The front and back cones of region $1$ add up to $90+90=180^\circ$, while regions $2$ and $3$ are $90^\circ$ each.

{\bf Voice source and mixture dataset:}
We use the LibriMix dataset \cite{librimix}, sampled at $16\textit{KHz}$, without considering noise and reverberation. 
With the script used in \cite{hungarian}, Libri5Mix is used for training and validation, while Libri2Mix, Libri3Mix, Libri4Mix, and Libri5Mix are used for testing.

% {\bf HRTF dataset:} For supervised region-based separation, we use the CIPIC HRTF database \cite{CIPIC} and the LibriMix audio dataset \cite{librimix}. 
% The CIPIC HRTF database contains real-world recorded Head Related Impulse Responses (HRIR)\footnote{HRTF is the frequency domain representation of HRIRs.} for $45$ subjects, with $25$ different azimuths and $50$ different elevations, at roughly $5$ degrees of angular resolution. 
% For our experiments, we only use the horizontal plane with $25$ different azimuth angles, divided into three regions, as shown in Figure \ref{fig:region}. 
% The front and back cones of region $1$ add up to $90+90=180^\circ$, while regions $2$ and $3$ are $90^\circ$ each.
% % The first region is the front-back region. 
% % Region 2 and Region 3 are left and right region, each with an angular span of 90 degrees.

% We use the LibriMix dataset (sampled at $16KHz$) without considering noise and reverberation. 
% With the script used in \cite{hungarian}, Libri5Mix is used for training and validation, while Libri2Mix, Libri3Mix, Libri4Mix, and Libri5Mix are used for testing. 

{\bf Creating binaural mixtures:}
To form a binaural mixture, we first assign a voice source to a randomly chosen region, and then select a random angle $\theta$ from within that region.
The voice source is then convolved with the corresponding HRIR($\theta$) -- this forms one of the components of the mixture.
To create a mixture of $K$ sources, we randomly choose $K$ from $[2,5]$, and repeat the same procedure.
With $K$ HRIR-convolved sources, we sum them to generate the mixture.
Observe that the HRIR is distinct for left and right ears, so we obtain a pair of mixtures -- called the binaural mixture.

% At training time, we randomly select between $2,3,4,5$ sources for each training mixture sample to decide how many voice sources we're actually mixing from the five sources in Libri5Mix. 

% For each data sample with several speeches, we mix them using the HRIRs. For each voice source, we first randomly select a region, and then from that region, we randomly select an HRIR. 
% The selected HRIR is then convolved with the voice source to be a component of the binaural mixture. 
% At training time, we randomly select from 2,3,4,5 for each training sample to decide how many voice sources we're actually mixing from the five sources in Libri5Mix. 

{\bf Generic vs. personalized training:}
To characterize the gap between generic and personal HRTF, we train two models:
(1) The training sources are all convolved with the test subject's personal HRIR -- as discussed earlier, this gives the upperbound on performance. 
(2) For the generic model, the training sources are convolved with a random person's HRIR, chosen randomly from a database of $27$ people's HRIRs.
% (2) For each training sample, one person is randomly selected from the HRTF database with 27 people. That person's HRIR is used for
During test, the model is tested with the test subject's HRIR. 

% (2) \hl{The training sources are convolved with different people's HRIRs, each HRIR drawn randomly from a pool of $27$ HRIRs.
% During test, the model is tested with a single subject's HRIR. }

% For the supervised training, we consider two cases. 
% The first case is to assume that we know the testing subject's HRIRs and thus we can directly train with those HRIRs. 
% In this case, we directly use the test subject's HRIR to do the mixing. This experiment's goal is to see what the model's upperbound is if the model is able to fit to only one single person.

% The second case is to train a general model using different people's HRTFs to make the model generalize all people's HRTFs. 
% In this case, we select 27 subject's HRTFs and for each data sample, we just randomly select a subject to use its HRIR for mixing. 
% During test, the model is still tested on single subject's HRIR. 
% For both cases, there are three sets of testing data, corresponding to three subjects.

{\bf Total users and models:}
For both cases, there are $3$ sets of testing data, corresponding to $3$ subjects.
Thus, overall we have $4$ supervised training models, i.e., $1$ generic HRTF-based model, and $3$ personalized models for the $3$ testing sets (from each subject).

{\bf Basic metric:} 
We use signal-to-noise-ratio (SNR) to assess separation quality.
Observe that our model outputs binaural sounds which needs to preserve the ILD, hence the commonly used SI-SDR metric \cite{roux2018sdr} does not apply.  
Thus, we compute SNR between a reference signal $x\in{R^{1\times{T}}}$ and the estimated signal $\hat{x}\in{R^{1\times{T}}}$ as: 
$$\text{SNR}(x, \hat{x})=10log_{10}(\frac{{{||x||}^2_2}}{{||x-\hat{x}||}^2_2})$$
{\bf Extending metric to region-wise mixtures:} 
For region-based separation, the notion of a voice source gets extended to a region-wise mixture.
So the reference signal in this case is the true mixture from that region, while the estimated signal is the estimated mixture from the same region.
In the special case where all sources are from the same region (and the other regions have no active sources), we simply use the same SNR equation from above --- we term this single-region SNR or ``S-SNR''.
However, when multiple regions are active, we modify the metric to SNR improvement (SNRi).
Assuming that the mixture of region-wise mixtures is denoted as $m\in{R^{1\times{T}}}$, we define SNRi as:
$$\text{SNRi}(x, \hat{x}, m)=\text{SNR}(x, \hat{x}) - \text{SNR}(x, m)$$
When $2$ regions are active, we average the two SNRi and report them as 2-SNRi. 
Since this is a binaural estimation, we average over the left/right microphones as well.
Similarly, for samples from $3$ active regions, we report 3-SNRi.

{\bf Results:}
Table \ref{table:self-supervision} reports the main performance results averaged over $3$ testing subjects for LibriKMix, where K is the number of sources between $2$ and $5$.
All results are in $dB$ scale and the model names are summarized below:

- {\bf \texttt{GENERAL:}} training with generic HRTFs. \\
- {\bf \texttt{PERSONALIZED:}} trained with subject's personal HRTF. \\
- {\bf \texttt{SELF:}} self-supervised training from stage 1 outputs. \\
- {\bf \texttt{SEMI:}} semi-supervised training with some clean sources. \\
- {\bf \texttt{FEW-SEMI:}} semi-supervised with less than $1hr$ dataset.

Note that \texttt{GENERAL} and \texttt{PERSONALIZED} are supervised, while \texttt{SELF}, \texttt{SEMI}, and \texttt{FEW-SEMI} are self-supervised. 
{\em Evident from the table, the SNR gap between \texttt{GENERAL} and \texttt{PERSONALIZED} is significant, characterizing the room for improvement available to self-supervised and personalization-based approaches.}

%Table caption
% Region-based model separation performance for supervised, self-supervised, and semi-supervised training. Supervised training includes general training and personalized training. All the results are the average of the three testing subjects. S-SNR means single-region SNR, 2-SNRi means two active region's average SNRi, 3-SNRi means three active region's average SNRi. Everything is in dB scale. Details is shown in section 4.1. General means supervised training using HRTF databases. Personalized means supervised training using test subject's HRTF. SELF mean self-supervised training, SEMI mean semi-supervsied training, and FEW-SEMI means smi-supervised training with dataset less than an hour.

\begin{table}[hbt]
\caption{Separation SNR compared between supervised and self-supervised models for increasing number of sources. The * indicates upper-bound performance with personal, clean sources.}
% \label{sample-table}
% \vskip 0.15in
\begin{center}
\begin{small}
\begin{sc}
\begin{tabular}{lccccr}
\toprule
Model & K & S-SNR & 2-SNR$i$& 3-SNR$i$ \\
\midrule
General    & 2& 21.0& 12.0& N/A\\
Personalized* & 2& 36.5& 16.7& N/A\\
Self    & 2& 31.4& 13.9& N/A\\
Semi    & 2& 33.1& 15.1& N/A\\
Few-Semi    & 2& 31.0& 15.0& N/A\\
\midrule

General    & 3& 20.9& 11.2 &13.1 \\
Personalized* & 3& 36.5& 15.3 &16.8 \\
Self    & 3& 32.2& 12.8 &14.9 \\
Semi    & 3& 33.5& 14.0 &15.7 \\
Few-Semi   &3 & 31.1& 13.8& 15.5 \\
\midrule

General    & 4& 20.5& 10.9 & 12.7 \\
Personalized* & 4& 36.3& 14.6 & 16.0 \\
Self    & 4& 33.9& 12.5 & 14.4 \\
Semi    & 4& 33.8& 13.4 & 15.0 \\
Few-Semi    & 4& 31.0& 13.2 & 14.8 \\
\midrule

General    & 5& 21.4& 10.2 & 12.2 \\
Personalized* & 5& 35.8& 13.7& 15.2\\
Self    & 5& 34.3& 11.8& 13.8\\
Semi    & 5& 33.8& 12.6& 14.3\\
Few-Semi    & 5& 31.0& 12.4& 14.1\\
\bottomrule
\label{table:self-supervision}
\end{tabular}
\end{sc}
\end{small}
\end{center}
\vskip -0.2in
\end{table}

\subsection{Self-supervised Training}
{\bf Creating the ``dirty'' source dataset:} 
For fair comparison, the training data for our self-supervised model is drawn from the same audio/HRTF dataset, except that they are deliberately mixed with interfering sources (other binaural speeches) and then fed to our {\em Selective Spatial Filter} in stage 1.
The output of stage 1, which still contains interference (hence called a ``dirty'' signal) is then used as training and reference signals in stage 2.
% The ``\hl{dirty}'' output is then used as training and reference signals in stage 2. 
Specifically, for each clean source $s^{r,l}$ in Libri5Mix, we convolve with a randomly chosen HRIR($\theta$) to create a binaural source.
Then we mix this binaural source with other sources that are also convolved with the same HRIR but a different random angle $\omega$. 
Of course, the mixing is done per channel to yield binaural mixtures.
When these mixtures are fed to our stage 1, the ``dirty'' output $\Tilde{s}^{l,r}$ serves as our self-training data.

% For the self-supervised training, for a fair comparison, we still use the same training data as the supervised training except we change the input mixtures and the reference region mixtures to be mixed from the separation result of selective spatial separation model. 
% For each clean source pairs $s^{r,l}$ in each sample of the training set of Libri5Mix, we generate a real-world separated version $\Tilde{s}^{l,r}$. 
% The way we do this is to first randomly select a speech sample from the LibriSpeech database and also one direction's HRIR to create a binaural mixtures by mixing $s^{l,r}$ and the randomly selected binaural sound, channel wise. 
% The way we select the HRIR for the randomly selected sound is to randomly sample HRIRs by making sure the ITD difference of the two sources between the two sources is bigger than $\Delta{\tau_{min}} = 0.0006$(second), which is about an average person's ITD difference when one sound is coming from the front and the other is coming from the right. 
% Then we feed this mixture to our selectively spatial clustering model to get the $\Tilde{s}^{l,r}$.

{\bf Configuring stage 1 model:} 
For stage 1's spatial clustering model, we use $1024$-point STFT with $512$ overlap using Hanning window. 
We set $f_{aliasing}$ = $562Hz$, which is about the $36^{th}$ bin in the FFT.
We set $\alpha = 5$, $\sigma_{th}=0.00007$ second --- this value was set empirically based on our discussion on Figure \ref{fig:sigma}.
% to ensure mixtures are not aggressively discarded.

\begin{figure}[t]
  \centering
  \includegraphics[width=3.3in]{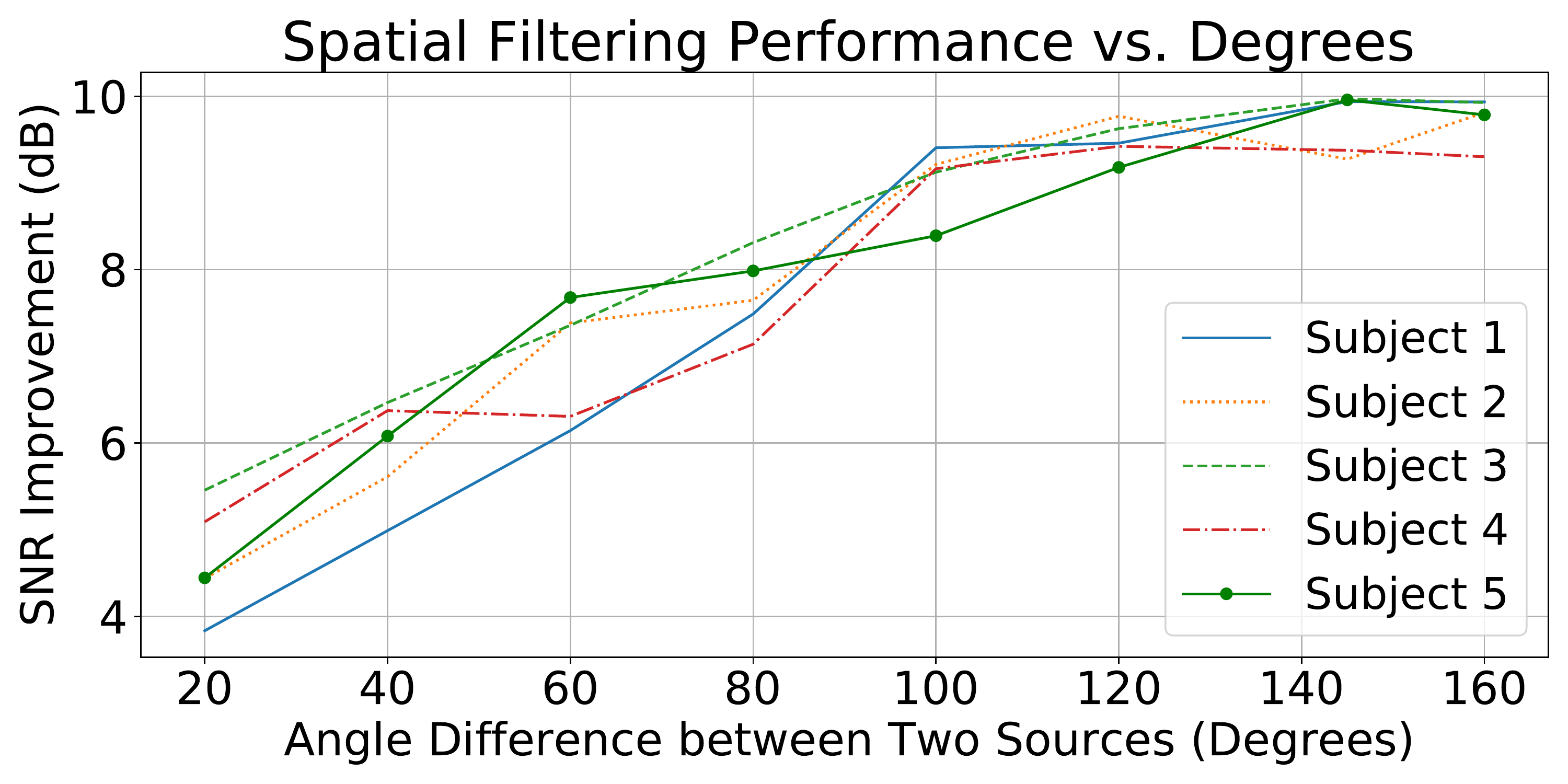}
  \vspace{-0.1in}
  \caption{Stage 1's ability to separate 2-source mixtures as a function of the angular separation between sources.}
  \vspace{-0.15in}
  \label{fig:spatial}
\end{figure}

{\bf Result:}
Figure \ref{fig:spatial} shows the performance of our spatial clustering algorithm for $5$ subject's HRTFs.
The separation improves as the angular separation increases between the sources in the mixture (recall that the filter only accepts $1$ or $2$ sources and discards $3+$ source mixtures).
% The testing data is Libri2Mix and figure shows the SNR improvement with respect to two sources' angle difference. 
% We show this performance for 5 subject's HRTFs.

{\bf Creating ``dirty'' mixtures:} 
Using the dirty sources $\Tilde{s}^{l,r}$ we generate the dirty mixtures $\Tilde{y}^{l,r}, \Tilde{m}^{l, r}$. 
While supervised training trains on $y^{l,r}$, self-supervised training --- using only 2-source mixtures --- trains on $\Tilde{y}^{l,r}, \Tilde{m}^{l, r}$. 
The performance is tested on many sources (Libri2Mix, Libri3Mix, Libri4Mix, and Libri5Mix).

{\bf Results:} 
Table \ref{table:self-supervision} shows the results.
Evidently, even if the training references are ``dirty'', they can still guide the model to outperform the general supervised model (trained on clean sources).
\texttt{SELF} outperforms \texttt{GENERAL} by $10$+$dB$ in terms of S-SNR and more than $1.5dB$ in terms of 2-SNRi and 3-SNRi.
Further, the performance improvement is higher with more sources.

% Then, the same as section 2.1, we can generate training input-output pairs $\Tilde{y}^{l,r}, \Tilde{m}^{l, r}$ from $\Tilde{s^{l,r}}$. 
% While supervised training trains on $y^{l,r}$, self-supervised training using only 2-source mixtures can train on $\Tilde{y}^{l,r}, \Tilde{m}^{l, r}$. 
% The performance tested on Libri2Mix, Libri3Mix, Libri4Mix, and Libri5Mix is shown in Table \ref{table:self-supervision}. 
% This result shows even if the training references are not very clean, it can still guide training to have much better performance than the generalized model.

\subsection{Semi-supervised and Few-shot training}
{\bf Semi-supervised} (\texttt{SEMI}):
Our dirty sources were all derived from mixtures.
In practice, when a user wears her earphone/hearing-aid or glasses in everyday life, their earphone is likely to record single sources as well --- in fact, they should be quite common.
% We also consider the case when we have some real world recordings those only contain one single source. 
% These recordings should be quite common in real world. 
To avail this benefit, we create the training samples with $50\%$ of clean sources and apply the same method for mixture generation.
We call this the semi-supervised model (\texttt{SEMI}) and add to Table \ref{table:self-supervision} as another point of comparison.

% To further achieve this, we generate the training input-output pairs $\Tilde{y}^{l,r}, \Tilde{m}^{l, r}$, in a way that for each clean source, we randomly decide if we mix using $\Tilde{s^{l,r}}$ or $s^{l,r}$. 
% We call the model trained this way semi-supervised trained model and the performance is also in Table \ref{table:self-supervision}.

{\bf Few-shot} (\texttt{FEW-SEMI}):
We further consider the case when there are limited amount of real-world recordings. 
Given that Libri5Mix offers $56$ hours of training mixtures, such a dataset may consume several days if a user must collect them in the real-world.
Thus, we only use $56$ minutes of personal data to fine-tune the \texttt{GENERAL} model, and then test if personalization can still outperform the $56$-hour trained generic-HRTF model.
% In the training set of Libri5Mix, there are 56 hours of training mixtures, which takes a long time to collect in real world, especially for personal recordings. 
% Thus, we consider only using 56 minutes of data to fine tune based on the general model. 

{\bf Results:}
Table \ref{table:self-supervision} shows favorable results, indicating that \texttt{FEW-SEMI} can learn the personal HRTF even from limited data, thus preserving most of the gains over \texttt{GENERAL}. Further, the small gap between \texttt{FEW-SEMI} and \texttt{SEMI} suggests that region-based separation generalizes well (instead of over-fitting).

%

% The result shows that, even with limit amount of real-world data, the fine tuning is able to learn the personalized model very well. 
% The performance of this is shown in Table \ref{table:self-supervision}.

% \begin{table}[H]
% \caption{Few shot semi-supervised training result}
% \label{sample-table}
% \vskip 0.15in
% \begin{center}
% \begin{small}
% \begin{sc}
% \begin{tabular}{lccccr}
% \toprule
% Model & N & SS & 2SI & 3SI \\
% \midrule
% General    & 2& 96.7$\pm$ 0.2& $\surd$ \\
% Personalized & 3& 80.0$\pm$ 0.6& $\times$\\
% Self-Supervised    & 4& 83.8$\pm$ 0.7& $\surd$ \\
% Semi-Supervised    & 5& 78.3$\pm$ 0.6&         \\

% \label{table:fewshot}

% \end{tabular}
% \end{sc}
% \end{small}
% \end{center}
% \vskip -0.1in
% \end{table}

\subsection{Classical Vs. Region-based Source Separation}
% Region-based Separation Versus Source Separation Using PIT
{\bf New metric:}
It is difficult to compare region-based separation with classical each-source-separation using permutation invariant training (PIT).
Hence we consider one special case of target speech extraction where the target speaker is alone in the front-back region and all other speakers (interferers) are in other regions. 
The goal is to compare the SNR of only the separated target speech.

{\bf Dataset:}
We train the neural network with identical settings to perform classical source separation using PIT with SNR loss. 
We train on Libri4Mix, and Libri5Mix separately, to obtain $2$ models. 
{\em This means the classical source separation model is assuming the number of sources is known, while the proposed  region-based model does not.}
We use Libri4Mix, and Libri5Mix as the test set for this experiment.
To synthesize binaural sounds, we use a randomly selected HRIR from region 1 (front-back region) for the target voice, and then randomly select HRIRs from other two regions for the interfering voices.

\begin{figure}[!h]
  \centering
\vspace{-0.1in}
  \includegraphics[width=3.2in]{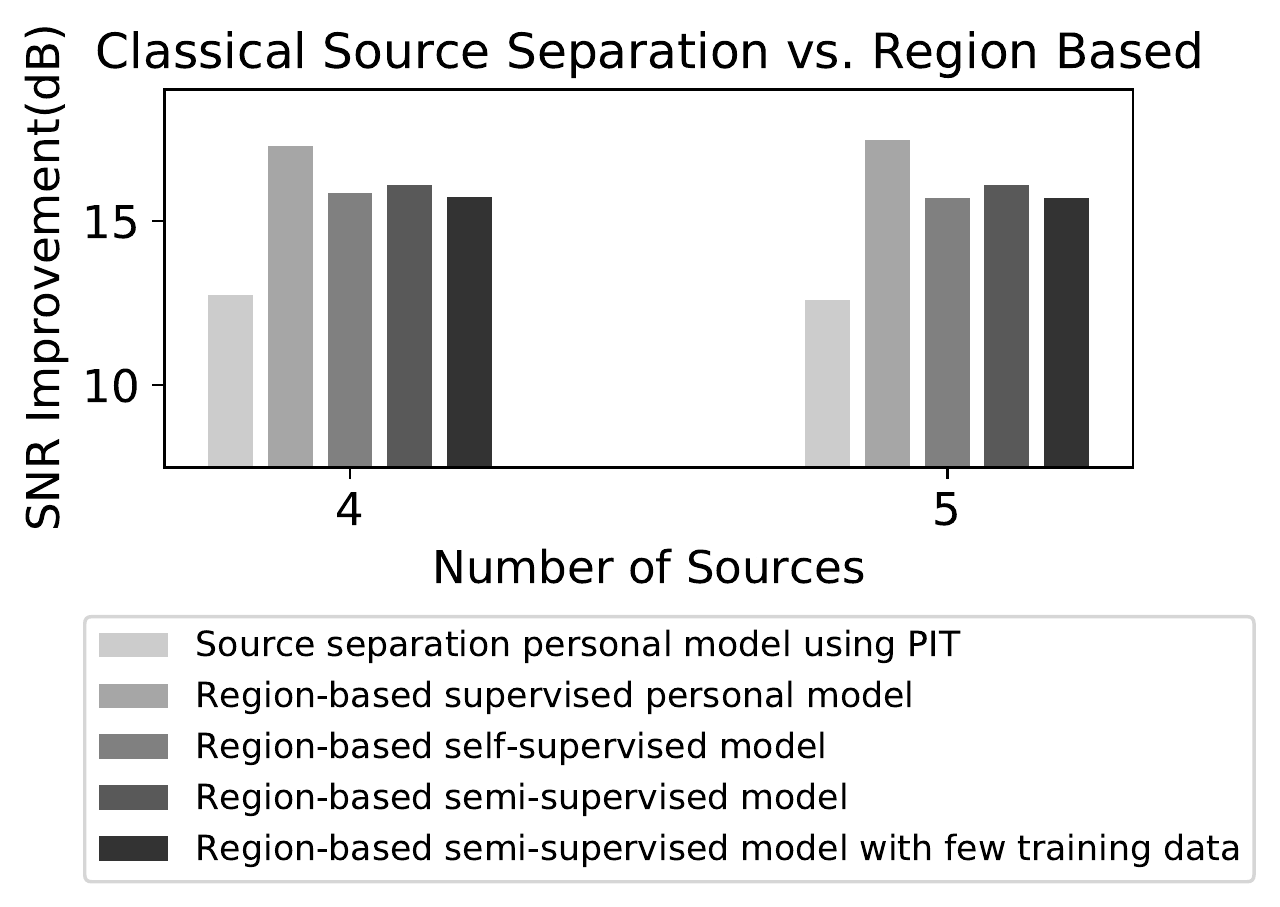}
  \vspace{-0.15in}
  \caption{Classical versus region-based separation: The target speaker is in region 1 while all interference speakers are in other regions. SNR improvement reported for target speech extraction.}
  \vspace{-0.1in}
  \label{fig:bar}
\end{figure}

{\bf Results:}
Figure \ref{fig:bar} plots the results.
Evidently, even though the classical source separation model assumes the correct number of sources, its performance for separating the target speech is substantively worse than the proposed region-based model. 
This result is strongly suggestive that spatial information may be far more valuable than spectral information, when it comes to separating multi-channel binaural voice recordings.

% \subsection{Classical Model Performances}
% classical model's performance:
%     1. single source variance: graphs: subject versus mean+-val for single source, and double sources with 10 degrees and 5 degrees apart
%     2. double sources' mean and variacnes: graphs: ITDs versus separation result, variances versus subjects
% TCN model's performance:
%     supervised versus self-supervised versus semi-supervised versus few-shot semi-supervised with 2,3,4,5 speakers

% \subsection{Separation Network}
% The deep learning model we use for region-based binaural voice separation is the interchannel feature concatenation TasNet in \cite{binaural}. The input of the model is the binaural mixture recordings, and the output of the model is the binaural sound for each region. Note that the sound we output should still preserve the spatial cues, which might be useful for later processing. $sin(ITDs)$, $cos(ITDs)$, and $20log_{10}(ILDs)$ are used as interchannel features. These features are concatenated with the mixture encoding for the TCN separator to generate the masks.

\subsection{Points of Discussion}
\begin{figure}[!h]
% \vspace{-0.05in}
  \centering
\includegraphics[width=3in]{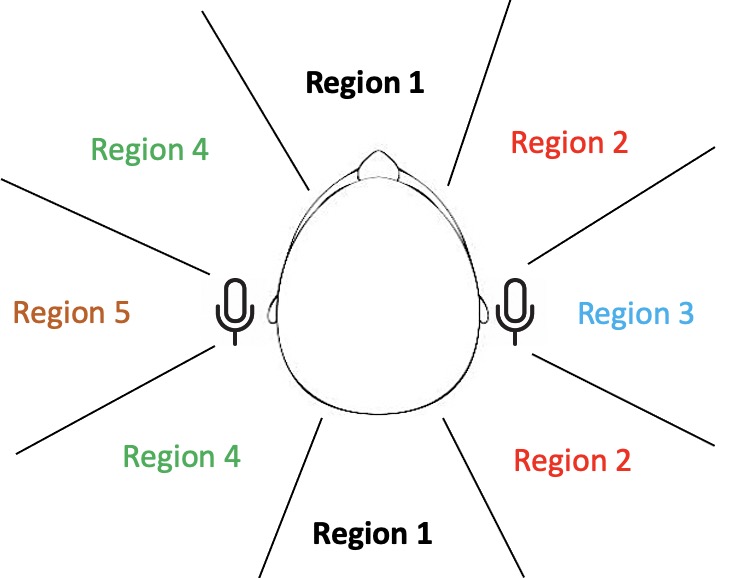}
  \vspace{-0.1in}
  \caption{Example of $5$ regions such that sources with same ITDs fall in the same region.}
\vspace{-0.2in}
  \label{fig:5regions}
\end{figure}

{\em Why not increase the number of regions?}
Recall from Figure \ref{fig:region} that the front and back cones together form a single region (Region $1$).
This is because we want signals producing the same ITDs to be located in the same region.
It is possible to increase the number of regions while still satisfying this property --- Figure \ref{fig:5regions} shows an example with $5$ regions.
However, such designs are not free of tradeoffs.
Specifically, at $16kHz$, the typical time difference of arrival (TDoA) is around $0.3ms$ which translates to $5$ audio samples. 
This means $5$ samples need to embed the spatial signature of any given region, a fundamentally difficult proposition even for deep neural networks. 
As compute power and number of microphones increase in earables, separating voices into more regions will become easier.
% Moreover, most of the audio separation applications are real-time in nature, so even if sampling rates are increased, the real-time deadlines are hard to meet at such high sampling rate.

% The first region is the front-back region. 
% Region 2 and Region 3 are left and right region, each with an angular span of 90 degrees.

{\em For our region setup, what if a source lies near the boundary between two regions?} To
cope with this, it is possible to define soft region boundaries, i.e., assign the source to the neighbor region if the source is very close to the boundary ($\approx 10^{\circ}$). However, in our experiments, we did not need to tackle this issue because we
were limited by the resolution of the HRTF database. 
In other words, since HRTFs are available at a granularity of $10^{\circ}$, we assigned nearby sources to the closest HRTF angle, which automatically made the region assignment.

\section{Related Work}

{\bf Single-Channel Speech Separation.}
Single channel source separation methods like \cite{2019,luo2020dualpath,subakan2021attention} are able to separate speech sources successfully using only single channel input. These models are all trained with permutation invariant training\cite{yu2017permutation, uPIT}. Deep clustering \cite{hershey2015deep, 8462507, attractor1, attractor2} is another approach to the permutation problem. Current solutions to the unknown number of sources problem are from 2 strategies. \cite{DBLP:journals/corr/abs-2011-02329,nachmani2020voice} tries to solve the problem of variable sources by assuming a maximum $K$. \cite{takahashi2019recursive} tries to solve this problem by decoding sources in a recursive manner until no sources are left.

{\bf Neural Binaural Speech Separation.}
Binaural recordings have also been used for neural speech separation. With multiple microphones, spatial information offers another cue for source separation. \cite{gu2019endtoend, gu2020enhancing} tries to learn inter-channel features for multi-channel speech separation. \cite{binaural,sagrnn} uses parallel shared encoders for binaural speech separation, while preserving the interaural cues. \cite{jenrungrot2020cone} proposes a binary search algorithm to continue searching for active sources. This work is similar to ours in the sense that it also uses region-wise separation to solve the unknown $K$ problem. Since they use $4$ microphones in their experiments, their model can localize sound sources with fine granularity. However, in the binaural case (e.g., earphones and hearing aids) front back confusion limits the approaches in literature.

{\bf Classical Binaural Speech Separation.}
Binaural speech separation without neural models is a well studied topic. 
These methods are essentially aims to cluster the T-F bins of the mixture based on interaural cues.
DUET\cite{duet} clusters using $2$ microphone recordings assuming no spatial aliasing.
EM based methods\cite{mandel2007algorithm, mandel2007localization, mandel2008, mandel2009model} attempt to exploit binaural cues like interaural time difference(ITD) and Interaural level difference(ILD) to cluster the T-F bins in STFT.
To avoid spatial aliasing, they employ graphical models to model each bin's ILD and IPD distributions.
These methods also assume approximate W-disjoint Orthogonality\cite{duet} which are violated with many source mixtures.
However, these methods achieve reliable performance with few sources, especially when they are not close to each other. 

{\bf Self-Supervised Neural Speech Separation.}
\cite{RIR_prob} shows that supervised speech separation model's performance degrades when channels mismatch between training and testing data. Certain self-supervised and unsupervised models are specifically designed to mitigate this problem. \cite{drude2019unsupervised, 2019unsupervised_2, seetharaman2018bootstrapping} uses spatial clustering to guide deep clustering.
% \hl{However, they do not consider binaural recordings.}
% assume the cases of microphones arrays' recordings which is different from the binaural recordings. 
A limitation with these methods is that spatial clustering might generate clusters that contain several very close sources, which cannot guide source separation models to separate all the sources.
\vspace{-0.1in} 

\section{Conclusion}
\vspace{-0.05in}
The importance of spatial cues in voice separation has been studied extensively.
However, the gap between generic and personalized spatial cues has been relatively less explored.
This paper finds that the human's head-related filter embeds valuable spatial signatures that can be learnt at coarse granularity (i.e., region-wise).
The performance gains are robust, and importantly, can be achieved in a self-supervised manner. 
Moreover, such region-wise voice separation also obviates the need to know the number of sources, thus relaxing an important assumption in practice.
We believe the findings could aid important applications for hearing aids and  earphones, such as selective hearing, noise cancellation, and audio-based augmented reality.

\section{Acknowledgments}
We thank NSF (award numbers: 1918531, 1910933, 1909568,
and 2008338, MRI-2018966) and NIH (award number: 1R34DA050262-01) for partially funding this research.

\label{submission}

% Planning:

% 0/ Bring introduction from previous report. \\
% 1/ Bring text from previous report into this paper. \\
% 2/ Revise related work and formulation to make it formal. \\
% ---- do the above by Wednesday -----

% ---- Finish experiments by Friday night and have results ready ---- \\

% 3/ Write/revise model and algorithm section: write carefully by breaking into appropriate subsections (e.g., training, two-stage, etc.) make it formal. \\
% 4/ Write dataset and evaluation section. \\
% 5/ Write Conclusion section.
% ---- do the above by Saturday -----------

\clearpage 

\nocite{langley00}

\bibliography{camera_ready}
\bibliographystyle{icml2022}

%%%%%%%%%%%%%%%%%%%%%%%%%%%%%%%%%%%%%%%%%%%%%%%%%%%%%%%%%%%%%%%%%%%%%%%%%%%%%%%
%%%%%%%%%%%%%%%%%%%%%%%%%%%%%%%%%%%%%%%%%%%%%%%%%%%%%%%%%%%%%%%%%%%%%%%%%%%%%%%
% APPENDIX
%%%%%%%%%%%%%%%%%%%%%%%%%%%%%%%%%%%%%%%%%%%%%%%%%%%%%%%%%%%%%%%%%%%%%%%%%%%%%%%
%%%%%%%%%%%%%%%%%%%%%%%%%%%%%%%%%%%%%%%%%%%%%%%%%%%%%%%%%%%%%%%%%%%%%%%%%%%%%%%
\newpage
\appendix
\onecolumn

% You can have as much text here as you want. The main body must be at most $8$ pages long.
% For the final version, one more page can be added.
% If you want, you can use an appendix like this one, even using the one-column format.
%%%%%%%%%%%%%%%%%%%%%%%%%%%%%%%%%%%%%%%%%%%%%%%%%%%%%%%%%%%%%%%%%%%%%%%%%%%%%%%
%%%%%%%%%%%%%%%%%%%%%%%%%%%%%%%%%%%%%%%%%%%%%%%%%%%%%%%%%%%%%%%%%%%%%%%%%%%%%%%

\end{document}